\documentclass{ws-procs975x65}

\usepackage{amssymb}

\begin{document}

\title{
Black Holes in Higher Dimensions\\ (Black Strings and Black Rings)
}

\author{Burkhard Kleihaus and Jutta Kunz}

\address{Institut f\"ur Physik, Universit\"at Oldenburg, Postfach 2503\\
D-26111 Oldenburg, Germany\\
E-mail: b.kleihaus@uni-oldenburg.de; jutta.kunz@uni-oldenburg.de}

\begin{abstract}
The last three years have again seen new exciting developments in the area
of higher dimensional black objects.
For black objects with noncompact higher dimensions, 
the solution space was exlored further within the blackfold approach
and with numerical schemes, yielding a large variety of new families of 
solutions, while limiting procedures created 
so-called super-entropic black holes.  
Concerning compact extra dimensions, 
the sequences of static nonuniform black strings
in five and six dimensions were extended to impressively large values of the
nonuniformity parameter with extreme numerical precision, showing that
an oscillating pattern arises for the mass, the area or the temperature,
while approaching the conjectured double-cone merger solution.
Besides the presentation of interesting new types of 
higherdimensional solutions,
also their physical properties were addressed in this session.
While the main focus was on Einstein gravity, a significant number of talks
also covered Lovelock theories. 
\end{abstract}


\bodymatter

\section{Introduction}\label{intro}
Starting with Kaluza and Klein the presence of higher dimensions
has been a recurring theme in our quest for a unified theory of the
fundamental interactions, culminating in the formulation of string
theory with its additional dimensions needed for its consistency.
With respect to gravity alone one often considers the number of
spacetime dimensions $D$ as a parameter of the theory,
where the study of the dependence on this parameter
may lead to new insights into the theory.
Concerning black holes this study has indeed brought forward
many new and unexpected features, which could not have been
anticipated given our knowledge of black holes in four dimensions.

When a compact flat direction is added to a Schwarzschild black hole
a black string emerges, where the horizon completely wraps
the compact extra dimension. The horizon topology is then given
by $S^2 \times S^1$, where the $S^1$ represents the compact
dimension. When the length $L$ of the compact dimension is
large as compared to the Schwarzschild radius $r_{\rm H}$ of the
black hole, the black string is thin and unstable as demonstrated by
Gregory and Laflamme \cite{Gregory:1993vy} long ago. 
For a given radius  $r_{\rm H}$
there is a crititical length $L_{\rm cr}$, where the stability changes,
so a fat black string is stable. At the critical point a new branch of
black string solutions emerges, which are no longer uniform
w.r.t.~the compact coordinate
\cite{Gubser:2001ac,Wiseman:2002zc,Wiseman:2002ti,Kleihaus:2006ee}. 
This branch of nonuniform black
strings has now been extended \cite{Kalisch:2015via} 
far closer to the topology
changing merger point \cite{Kol:2002xz,Kol:2004ww}, 
where a branch of caged black holes
is encountered \cite{Kol:2003if,Sorkin:2003ka,Kudoh:2004hs,Headrick:2009pv}.
By adding more compact dimensions black branes are obtained,
which also suffer from the Gregory-Laflamme instability.

In the case of noncompact extra dimensions the Kerr solution was
generalized by Myers and Perry \cite{Myers:1986un}, 
obtaining rotating black holes
with spherical horizon topology. Interestingly,
in six and more dimennsions, there is no upper bound on the
angular momentum of the singly spinning Myers-Perry black holes
for a given mass.
This led Emparan and Myers \cite{Emparan:2003sy} to the conjecture,
that there should be a dynamical bound of the angular momentum.
Based on the geometry of the highly flattened horizon
they argued that these overrotating black holes should
develop Gregory-Laflamme type instabilities,
giving rise to new branches of black holes, whose
horizon would be deformed.
Indeed, such pinched black holes have been found recently
\cite{Dias:2014cia}
showing the consistency of the envisaged phase diagram 
\cite{Emparan:2007wm,Emparan:2010sx}.

In six dimensions this phase diagram includes black rings,
which come with two branches \cite{Kleihaus:2012xh},
a thin black ring branch
and a fat black ring branch, similar to the case of five dimensions.
However, here the fat black ring branch does not extend to
zero area, but is expected to end at a horizon topology changing
merger solution with a finite area, where it
merges with a branch of pinched black holes.
Also the phase diagram for a more general set of solutions
has been studied recently, where the black ring is generalized to incorporate
a horizon topology $S^2 \times S^3$, which has been dubbed black ringoid
\cite{Kleihaus:2014pha,Kleihaus:2015dna}.

While those ringlike solutions were obtained numerically, 
their existence had been predicted within the blackfold approach 
\cite{Emparan:2009cs,Emparan:2009at,Emparan:2009vd}.
This approach needs two different scales which in this context
are given by the large size of the ring and the small size of the
black hole horizon. Comparison of the numerical solutions
with the blackfold predictions showed, that the blackfold
remains reliable far longer than expected, giving still acceptable
predictions as the branch of fat black rings/fat ringoids is approached.
It finally starts to fail in the vicinity of the cusp, where the
thin and the fat black ring/black ringoid branches merge.

While the blackfold approach predicts a large number of new
solutions, there may be further solutions, which do not satisfy
its basic assumptions, but which might exist nevertheless
as part of the phase space of higherdimensional black objects.
A nice introduction to blackfolds and to some of its intriguing
beasts was the first highlight of the session.
Another set of intriguing black holes presented
possess at the same time a noncompact horizon,
which is topologically a higher-dimensional sphere with two punctures,
and a finite area. Violating the isoperimetric inequality conjecture,
these black holes have been named super-entropic
\cite{Gnecchi:2013mja,Klemm:2014rda,Hennigar:2014cfa,Hennigar:2015cja}.

But also seemingly boring higherdimensional black holes
with spherical horizon topology were recently observed to exhibit
fascinating new features \cite{Blazquez-Salcedo:2013muz}.
The prerequisite here was the presence of a gauge field
with a Chern-Simons term. When the Chern-Simons coupling
is sufficiently large, sequences of radially excited black holes
arise, there is violation of uniqueness between extremal
and nonextremal black holes, and near-horizon solutions
are seen to correspond to an infinite sequence of global
black holes, a single global black hole, or no black hole at all.

As discussed in detail in the following in section 2, the session covered
many new exciting results on black holes in Einstein gravity. Moreover,
Lovelock gravity and its interesting black holes were a second major
topic of the session, as discussed in section 3.

\section{Solutions in higherdimensional Einstein gravity}

In five spacetime dimensions
there are a number of construction methods to obtain
exact solutions. However, there does not
seem to exist an analytic framework in order to obtain
solutions in more than five dimensions.
Consequently, one has to either resort to perturbative methods
or to numerical methods at the moment.

In the following we will discuss the various types of new solutions
of higherdimensional black objects in Einstein gravity,
that were presented in the session.

\subsection{Solutions with nonspherical horizon topology}

Starting with black objects in noncompact extra dimensions
we will then turn to discuss black objects with compact extra dimensions.

\subsubsection{Solutions in the blackfold approach}

In their seminal paper \cite{Myers:1986un}
Myers and Perry suggested a
heuristic way to construct black rings:
take a black string, bend it, and spin it along
the ring $S^1$ direction to achieve balance
\cite{Emparan:2001wn,Emparan:2008eg}.
This heuristic picture forms the basis of
the perturbative technique of matched asymptotic expansions
\cite{Emparan:2007wm}
and the blackfold approach 
\cite{Emparan:2009vd,Emparan:2009at,Emparan:2009cs}.

The blackfold approach has been suggested to obtain perturbative solutions
for black objects in more than five dimensions,
when there are two scales associated with the solutions,
which are sufficiently far apart 
\cite{Emparan:2007wm,Emparan:2009vd,Emparan:2009at,Emparan:2009cs}.
An ab initio derivation of the 
blackfold effective theory has been given in \cite{Camps:2012hw}.

In his exciting talk on {\sl New Geometries for Black Hole Horizons: 
a Review of the Blackfold Approach} 
the invited speaker Jay Armas first explained the
basic principles of the blackfold approach.
In particular, he explained the building of the effective theory,
and he argued how the physical properties
of black objects can be captured by the physics of fluid flows
\cite{Armas:2011uf,Armas:2012jg}.

Subsequently, Jay Armas concentrated on the question of how to 
scan for horizons. Based on his results in \cite{Armas:2015kra}
he provided evidence for the existence of rather involved
horizon geometries and topologies, and he argued
that also plane wave spacetimes allow for a very
rich phase structure of higherdimensional black holes. 
As a key ingredient he employed
results from classical minimal surface theory.
He then explicitly constructed blackfold solutions 
consisting of planes, helicoids, catenoids and Scherk surfaces
\cite{Armas:2015kra}.

Next Jay Armas considered the elastic expansion 
\cite{Armas:2013hsa,Armas:2013goa}.
He pointed out that stationary fluid brane configurations 
are characterized by a set of elastic, hydrodynamic and spin 
response coefficients, and that the elastic 
and hydrodynamic degrees of freedom are coupled,
and he employed the second order effective action of stationary blackfolds 
to obtain the higher order corrected properties of thin black rings
\cite{Armas:2014bia}. Indeed, for black rings in seven dimensions,
the agreement of the corrected perturbative results 
with the exact numerical results \cite{Dias:2014cia} is most impressive.
The agreement remains excellent also in the presence of 
an electric charge, as shown for the case of
black rings carrying Maxwell charge in Einstein-Maxwell-
Dilaton theory with the Kaluza-Klein coupling constant
\cite{Armas:2014rva}.

In the last part of his talk Jay Armas
considered worldvolume effective actions for black holes 
by integrating out further scales
\cite{Armas:2015nea}.
In particular, he presented novel geometries for black hole horizons 
in higherdimensional asymptotically flat space-time, 
concentrating on helicoidal black rings in $D>5$
and helicoidal black tori in D = 7.

\subsubsection{Black ringoids}

While the blackfold approach is excellent and well justified
when the lengthscales in the problem are very different,
one has to resort to numerical methods, when the length
scales become comparable, in order to complete the 
phase diagram of higherdimensional black objects.

In his invited talk on {\sl Black Ringoids: New Higherdimensional
Black Objects with Nonspherical Horizon Topology}, Eugen Radu
started by recalling the well known phase diagram in five dimensions
for Myers-Perry black holes and black rings,
noting that both meet in a singular solution with vanishing
horizon area, and discussed the blackfold limit.

Subsequently, Eugen Radu described a general framework for
obtaining black holes with $S^p \times S^q$ horizon topology,
concentrating on the case of $S^{D-(2k+3)} \times S^{2k+1}$
topology, where the spinning part is
the $S^{2k+1}$ part \cite{Kleihaus:2014pha}.
For an appropriately chosen value of the horizon angular velocity,
the solutions become balanced.
A list of such solutions for $D \le 11$ is shown in Table 1.

\begin{table}[h!]
\centering
\begin{tabular}{|c|c|c|c|c|c|}

\hline
 $~$ & ${\it spherical~horizon}$ & {\it black~rings} & \multicolumn{3}{c|}{{\it black~ringoids}  }
\\
\hline
$~$ & MP/`pinched'
  &  $k=0$ &  $k=1$ &  $k=2$ &  $k=3$
\\
\hline
\hline
$D=5$   & $S^3$ &  ${\mathbf S^2}\times {\mathbf S^1}$ & $~$  & $~$ & $~$
\\
$D=6$   & $S^4$ & $S^3\times S^1$ & $~$  & $~$ & $~$
\\
$D=7$   & $S^5$ & $S^4\times S^1$ & ${\mathbf S^2}\times {\mathbf S^3}$  & $~$ & $~$
\\
$D=8$   & $S^6$ & $S^5\times S^1$ & $S^3\times S^3$  & $~$ & $~$
\\
$D=9$   & $S^7$ & $S^6\times S^1$ & $S^4\times S^3$  & ${\mathbf S^2}\times {\mathbf S^5}$   & $~$
\\
$D=10$  & $S^8$ & $S^7\times S^1$ & $S^5\times S^3$  & $S^3\times S^5$   & $~$
\\
$D=11$  & $S^9$ & $S^8\times S^1$ & $S^6\times S^3$  & $S^4\times S^5$ & ${\mathbf S^2}\times {\mathbf S^7}$
\\
 \hline
\end{tabular}
\label{table1}
\end{table}
\vspace*{-0.5cm}
{\small {\bf Table 1.} Horizon topologies for spinning balanced
black objects considered in \cite{Kleihaus:2014pha}.}
\vspace*{0.3cm}

Eugen Radu emphasized, that the solutions considered so far
had all equal angular momenta on the rotating $S^{2k+1}$,
the reason being, that in this case the problem simplifies to
a codimension-two problem. Moreover, he explained that
in order to construct such solutions numerically, it turned
out to be crucial to employ an adequate coordinate system.
Then the solutions could be obtained by solving the resulting set of PDEs 
subject to appropriate boundary conditions, where
a finite difference solver and a spectral solver were used,
with very good agreement between the results of the two different methods.
The appropriate value of the horizon angular velocity for balanced
solutions could be found by a shooting procedure.

Setting the integer $k=0$, one obtains higherdimensional black rings,
while the case $k \ge 1$ produces new nonperturbative black objects,
which Eugen Radu named black ringoids.
He then continued to first recall the solutions for balanced black rings 
in six dimensions, pointing out 
the very good agreement with the blackfold approach
on the one hand, and the new features 
of the black rings in $D \ge 6$ on the other hand \cite{Kleihaus:2012xh}.
In particular, for black rings in $D \ge 6$ the fat black ring
branch does not extend to vanishing area, but at a finite value
a horizon topology changing transition to 'pinched' Myers-Perry
black holes should occur \cite{Emparan:2003sy,Dias:2014cia}.
Thus there is a fundamental difference between the case
$D=5$, where the area at the merger point of black rings
and black holes vanishes, and the case $D \ge 6$.

As Eugen Radu explained, the phase diagrams of black holes and black rings
in five dimensions and in six dimensions represent generic situations, 
which are encountered also for the more general configurations,
the black ringoids \cite{Kleihaus:2014pha,Kleihaus:2015dna}.
In particular, for the $k=1$ black ringoids in seven dimensions
the pattern of the phase diagram of the black rings in five dimensions
is repeated, as can be concluded from the numerical results.
A theoretical argument here is based on the behavior of the
respective family of Myers-Perry solutions with two equal
angular momenta and the third angular momentum vanishing.
The angular momentum of this family of solutions does not 
extend arbitrarily far for a given mass, but a maximal value is reached
for a singular configuration with vanishing area.
On the other hand, in more than seven dimensions
the respective Myers-Perry black holes with two equal angular momenta
become overrotating and instabilities should arise,
associated with 'pinched' black holes
at a merger solution with finite area.
Therefore the branch of fat $k=1$ black ringoids in $D \ge 8$
should not extend to vanishing area, but a horizon topology
changing transition should be encountered
to a respective branch of 'pinched' black holes.

While an analogous set of arguments should hold for the $k \ge 2$ 
black ringoids,
the general picture Eugen Radu unveiled for these black objects
represents only the tip of the iceberg, with innumerous
further possibilities remaining to be studied.
Still, he expressed his hope, that one may find a
{\sl periodic table of black objects}, i.e., a classification
of all higherdimensional black objects based on a finite
number of simple features.

Eugen Radu concluded his talk with a discussion of 
static solutions with horizon topology $S^2 \times S^{D-4}$ obtained
previously,
which possess a conical singularity \cite{Kleihaus:2009wh,Kleihaus:2010pr}.
Allowing for further fields, in particular, a Maxwell field,
he then addressed static asymptotically flat magnetic configurations
\cite{Kleihaus:2013zpa},
which can be viewed as generalizations of the $D=5$ 
static dipole black ring \cite{Emparan:2004wy}.
They can be balanced by ``immersing'' them
in a background gauge field via a magnetic Harrison transformation
\cite{Ortaggio:2004kr,Kleihaus:2013zpa}.
Since the magnetic field does not vanish asymptotically,
the resulting configurations correspond to 
balanced black objects in a Melvin universe background
\cite{Kleihaus:2013zpa}.

\subsubsection{Super-entropic black holes}

An intriguing new type of solutions was presented by
David Kubiznak in his talk titled
{\sl Super-Entropic Black Holes}.
Starting out with a prelude on the horizon topology,
he first reminded the audience of Hawking's theorem,
requiring a spherical horizon topology 
for asymptotically flat black holes in four dimensions
\cite{Hawking:1971vc}.
He then turned to AdS black holes, where the horizons
may be either compact or noncompact,
and compact horizons may be Riemann surfaces of arbitrary genus $g$
\cite{Vanzo:1997gw},
while noncompact horizons may, for instance,
correspond to hyperboloid membranes \cite{Caldarelli:2008pz},
etc.

A very interesting new result here is the existence of black holes, 
which possess a noncompact horizon and at the same
time a finite horizon area and therefore finite entropy.
Such black holes were first discussed in 
\cite{Gnecchi:2013mja,Klemm:2014rda},
possessing horizons that are topologically
$(D-2)$-spheres with two punctures.
Thus there are even more surprizing event horizon topologies possible.

Since these black holes \cite{Gnecchi:2013mja,Klemm:2014rda}
may be viewed as a new type of ultraspinning 
limit of the Kerr-Newman-AdS solution,
David Kubiznak turned to the general discussion of
ultraspinning limits taken for Kerr-AdS black holes to
obtain new types of black holes \cite{Hennigar:2014cfa,Hennigar:2015cja}.

In the ultraspinning limit the rotation parameter $a$ tends to the
AdS radius $l$, yielding a singular limit, where three cases may
be considered.
i) In the brane limit (for $D>5$), one keeps the physical mass fixed
while simultaneously zooming in to the pole.
This corresponds to the AdS analogue of the Gregory-Laflamme type
instability of ultra-spinning Myers-Perry black holes 
\cite{Emparan:2003sy,Armas:2010hz}.
ii) In the hyperboloid membrane limit, the horizon radius is fixed,
while simultaneously zooming in to the pole
\cite{Caldarelli:2008pz},
yielding a rotating hyperboloid membrane with horizon topology
${\mathbb{H}} \times S^{D-4}$.

The so-called super-entropic limit iii) is obtained by 
first introducing a new azimuthal coordinate, by boosting 
the asymptotic rotation to the speed of light, and by
compactifying the azimuthal direction.
In four dimensions one then obtains a solution 
whose horizon is noncompact while being of finize size,
which possesses an ergosphere, and no obvious pathologies,
while the symmetry axis represents a boundary excised from the spacetime.

David Kubiznak then explained the name super-entropic.
Since in AdS black hole spacetimes one can identify the
negative cosmological constant with a positive pressure,
one can also define the thermodynamic volume of a black hole
\cite{Kastor:2009wy,Cvetic:2010jb}.
This led to the conjecture of the isoperimetric inequality,
which can be interpreted as the statement 
that the entropy inside a horizon of a given volume is
maximized for the Schwarzschild-AdS black hole
\cite{Cvetic:2010jb}.
Since the new solutions provide a non-compact counterexample
to this conjecture, they have been named super-entropic.

Subsequently, David Kubiznak applied the 
super-entropic limit analogously
to the rotating Myers-Perry-AdS solutions in $D$ dimensions
\cite{Hawking:1998kw,Gibbons:2004uw},
to obtain the corresponding higherdimensional type of solutions
\cite{Hennigar:2014cfa,Hennigar:2015cja}.
Here the entropy per given thermodynamic volume exceeds 
the limit set by the conjectured isoperimetric inequality,
at least in some parameter range, leading now to higherdimensional
``super-entropic'' black holes.

\subsubsection{Black strings}

Let us turn now to the case of compact extra dimensions
and focus on the new results on nonuniform black strings 
\cite{Kalisch:2015via}.
These were presented by Michael Kalisch in his talk on 
{\sl Highly Deformed Non-uniform Black Strings}.

In black strings the horizon completely wraps a compact extra dimension, 
with the horizon topology given by $S^{D-3} \times S^1$, 
where the $S^1$ now represents the compact dimension. 
While the branches of nonuniform black strings in five and six
dimensions were constructed before
\cite{Gubser:2001ac,Wiseman:2002zc,Wiseman:2002ti,Kleihaus:2006ee},
the new calculations extend very much further towards
the horizon topology changing merger \cite{Kol:2002xz,Kol:2004ww}
with a branch of caged black holes
\cite{Kol:2003if,Sorkin:2003ka,Kudoh:2004hs,Headrick:2009pv}.

\begin{figure}[t!]
\begin{center}
\includegraphics[height=.35\textheight, angle =0]{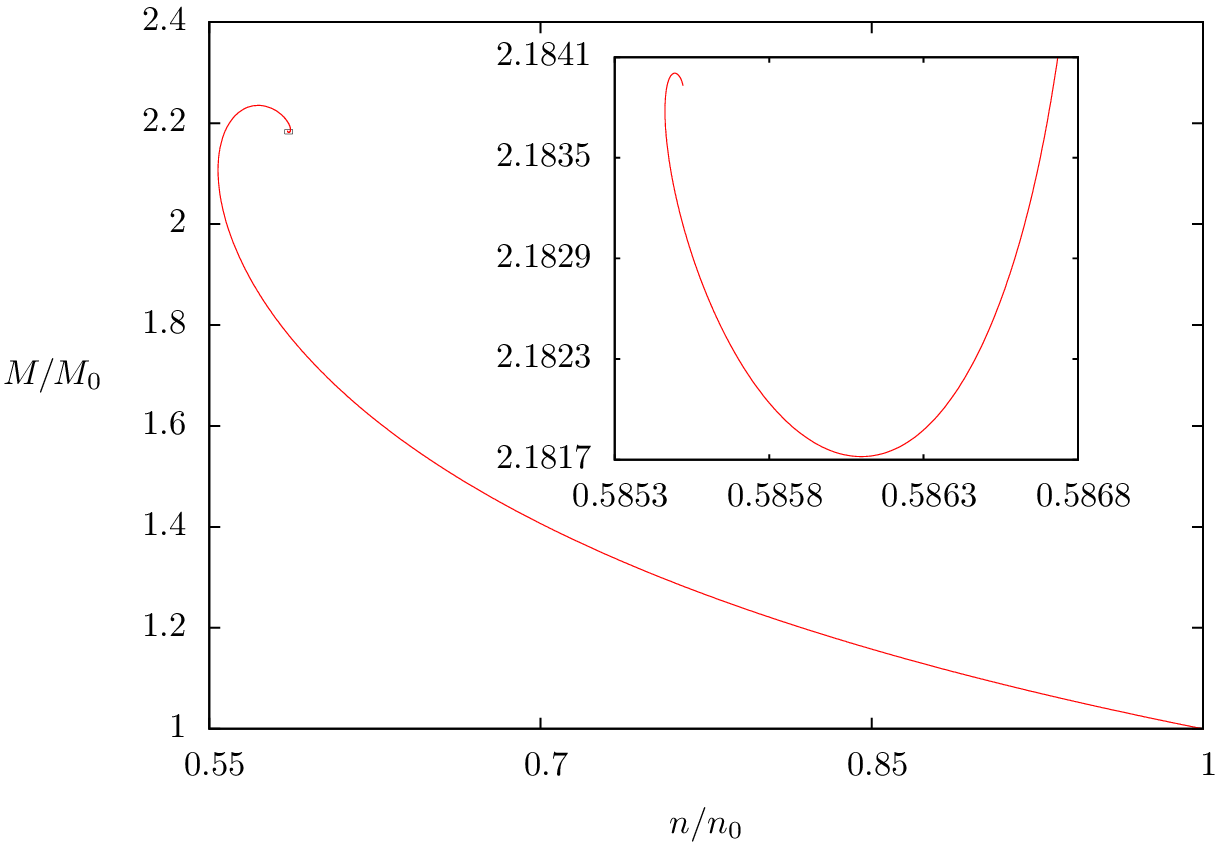}
\end{center}
\vspace{-0.5cm}
\caption{Spiral of the scaled mass versus the scaled relative tension
for nonuniform black strings in six dimensions
(courtesy to Michael Kalisch).}
\label{Fig1}
\end{figure}

The new results fully agree with the previous results in
\cite{Kleihaus:2006ee}, where a backbending of the branches
had been noticed, when the mass, area or temperature were
considered as functions of the relative tension.
When considered as a function of the nonuniformity parameter
$\lambda=\frac{1}{2}(R_{\rm max}-R_{\rm min})/R_{\rm min}$
this behavior is reflected in an oscillation of these
physical quantities.
While the previous calculations \cite{Kleihaus:2006ee}
are reliable up to $\lambda=9$, the new
calculations reach values of $\lambda$ of several hundred.

As Michael Kalisch explained, the numerical scheme to obtain
these highly accurate results is based on the use of a pseudo-spectral
method. Very important ingredients are furthermore a set of
appropriate coordinate transformations and the employment
of many (up to seven) domains. 
For $\lambda \approx 200$, for instance, he demonstrated 
a most impressive accuracy of the calculations, yielding
a value on the order of $10^{-13}$ for the residual
and for the deviation from the Smarr formula,
leaving the audience flabbergasted.

Thus with this method Michael Kalisch could not only
confirm the occurrence of a maximum in the mass,
but show two further extrema of the mass,
and likewise for the area, the temperature,
and the relative tension.
The new results therefore represent 
the beginning of a spiral, when the mass, the area,
and the temperature are considered versus 
the relative tension.
These exciting results are demonstrated in Figure 1
for the scaled mass versus the scaled relative tension
for nonuniform black strings in six dimensions
(where the scaling is w.r.t.~uniform black strings).

The calculations also strongly support the double-cone
structure of the merger solution \cite{Kol:2002xz,Kol:2004ww}.
To get a more complete understanding of the
black string -- caged black hole phase transitions,
it would be highly desirable, to employ analogous
sophisticated calculations also to the caged black hole
branch, to see, whether such a spiral-like behavior
arises there as well.
Moreover, it would be very interesting to
address with these superior numerical techniques also the 
horizon topology changing transition in more than six dimensions
\cite{Sorkin:2004qq,Sorkin:2006wp,Figueras:2012xj}.

\subsection{Solutions with spherical horizon topology}

Let us now turn to the seemingly simple higherdimensional
black holes with spherical horizon topology.
Also here interesting new developments have taken place
recently. In the session we have focussed
on the effect which matter distributions
and classical fields can have on the properties of 
such higherdimensional black holes.

\subsubsection{Black holes with distorted horizons}

Whereas the black holes discussed so far represent
isolated black holes, one may also consider black holes
which are not isolated but interacting with some external 
distribution of matter.
While this would lead to dynamical systems in the general case,
solvable only by numerical or perturbative methods,
exact solutions of non-isolated black holes
can be constructed as local solutions, physically relevant only in a
close neighborhood of the black hole, which still incorporate 
information on the external matter fields.

Such distorted black holes were first obtained by
Geroch and Hartle \cite{Geroch:1982bv} for the static case
in four dimensions, who also investigated various 
properties of these black holes.
Rotating distorted black holes were constructed and investigated
subsequently 
\cite{Tomimatsu:1984sx,Breton:1998sr,Breton:1997,Abdolrahimi:2015gea}.
As Petya Nedkova suggested in the introduction of her talk on 
{\sl Rotating Distorted Black Holes in Higher Dimensions},
such four-dimensional distorted black holes may, for instance, describe
black holes surrounded by an accretion disk.

Petya Nedkova then continued to describe the general procedure 
on how to obtain distorted black holes, starting with a
four-dimensional example. Expressing the metric in Weyl coordinates
\begin{equation}
d s^2 = - e^{2u} dt^2 + e^{-2u} \rho^2 d \phi^2 + 
e^{2(\gamma-u)} ( d \rho^2 + dz^2 )
\end{equation}
with metric functions $u(\rho,z)$ and $\gamma(\rho,z)$,
one finds the Laplace equation for $u$ in three-dimensional flat space
and a linear system for $\gamma$, which is always integrable
for a given solution $u$.
Since the uniqueness theorem does not extend to 
asymptotically non-flat vacuum solutions, one finds infinitely
many black holes with a regular horizon, i.e., distorted black holes.
The expansion coefficients in the solution for $u$ then
characterize the external matter distribution.

Turning to higherdimensional distorted black holes,
Petya Nedkova recalled the known static solutions
\cite{Abdolrahimi:2010xw,Abdolrahimi:2013cza},
and then addressed the construction of 
distorted Myers-Perry black holes with a single angular momentum
\cite{Abdolrahimi:2014qja}.
These solutions are obtained
by applying a two-fold B\"acklund transformation
on a five-dimensional vacuum Weyl solution as a seed, 
which describes a regular spacetime region in the presence 
of some external matter distribution.

This leads to five-dimensional stationary and axisymmetric
vacuum solutions with a single rotation parameter.
Their horizon has spherical topology, and the solutions
are asymptotically non-flat.
Therefore, Petya Nedkova interpreted the solutions
as describing locally a Myers-Perry black hole 
in the presence of an external matter distribution.
The solutions possess no curvature singularities outside
their horizon, and conical singularities can be avoided
by respecting a relation between the expansion coefficients
of the metric functions.
Petya Nedkova concluded her talk
with a discussion of the physical properties of these
distorted black hole solutions, considering in particular
the ergoregion for dipole and quadrupole distortions.

\subsubsection{Einstein-Maxwell black holes}

Francisco Navarro-L\'erida discussed in his talk the
{\sl Properties of Rotating Einstein-Maxwell-Dilaton 
Black Holes in Odd Dimensions}, restricting to asymptotic
flatness and a spherical horizon topology.
The general set of analytical Einstein-Maxwell-dilaton black holes
can be constructed by employing a Kaluza-Klein reduction
for a particular value of the dilaton coupling constant,
$h=h_{\rm KK}$
\cite{Chodos:1980df,Frolov:1987rj,Horne:1992zy,Kunz:2006jd}.
For different values of the dilaton coupling constant $h$, however,
one either has to employ perturbative techniques
\cite{Aliev:2004ec,Aliev:2005npa,Aliev:2006yk,NavarroLerida:2007ez,Sheykhi:2008bs,Allahverdizadeh:2010xx,Allahverdizadeh:2010fn}
or perform a numerical analysis
\cite{Kunz:2005nm,Kunz:2006eh,Blazquez-Salcedo:2013yba,Blazquez-Salcedo:2013wka},
while for extremal black holes one may, in addition,
resort to the near-horizon formalism to gain understanding
of the properties of these black holes 
\cite{Astefanesei:2006dd,Goldstein:2007km,Figueras:2008qh,Kunduri:2013gce}.

Specializing to black holes with equal angular momenta,
Francisco Navarro-L\'erida explained, 
that in odd-$D$ dimensions this leads to an enhancement of
the symmetry of the solutions,
and a substantial simplification of the equations to cohomogeneity-1,
such that only ODEs and not PDEs must be considered
\cite{Kunz:2006eh},
and that in the near-horizon formalism the resulting formulae
are all analytical.

The results presented for the near-horizon formalism were quite
intriguing. For the Einstein-Maxwell case ($h=0$), 
there are two sets of near-horizon solutions for all odd dimensions.
Denoting them as the Myers-Perry branch and the 
Reissner-Nordstr\"om branch,
since they emerge from the extremal Myers-Perry 
and Reissner-Nordstr\"om solutions, respectively,
Francisco Navarro-L\'erida pointed out,
that for the black holes on the Myers-Perry branch
the angular momenta are proportional to the horizon area,
while for the black holes on the  Reissner-Nordstr\"om branch
the angular momenta are proportional to the horizon angular momenta.
The two branches of near-horizon solution cross at a critical point.

Interestingly, the numerically obtained global solutions
do not exist for the full near-horizon branches.
Instead, only particular parts of these near-horizon branches 
are realized globally.
Thus the family of global solutions consists only of the
first part of the Myers-Perry branch and the
second part of the Reissner-Nordstr\"om branch,
meeting at the critical point.

In constrast,
for an arbitrary finite value of the dilaton coupling constant $h$,
there is only a single set of near-horizon solutions,
and the angular momenta are always proportional to the horizon area,
as Francisco Navarro-L\'erida explained.
He then addressed the domain of existence of 
Einstein-Maxwell-dilaton black holes, showing that 
it is determined by the set of static black holes
and the set of extremal rotating black holes.
Finally he noted, that one can infer
to good approximation the horizon angular velocity
of the extremal rotating black holes from the
surface gravity of the static black holes.

\subsubsection{Einstein-Maxwell-Chern-Simons black holes}

Even more surprizing results for 
extremal charged rotating black holes in higher dimensions
were reported by Jose Bl\'azquez-Salcedo in his talk 
{\sl Charged and Rotating Black Holes 
in 5D Einstein-Maxwell-Chern-Simons Theory}.
In odd spacetime dimensions one may add a Chern-Simons term to the
Einstein-Maxwell action. In the presence of this term
analytical black hole solutions were found
in five dimensions, when the Chern-Simons coupling constant
assumes a particular value $\lambda=\lambda_{\rm sg}$
corresponding to minimal supergravity
\cite{Breckenridge:1996is,Cvetic:2004hs,Chong:2005hr}.

For values of the Chern-Simons coupling constant $\lambda>\lambda_{\rm sg}$,
however, intriguing new behavior occurs
\cite{Kunz:2005ei,Kunz:2006yp,Blazquez-Salcedo:2013muz,Blazquez-Salcedo:2015kja}.
For instance, counterrotation sets in, where 
the horizon angular velocity and the angular momentum have opposite signs 
\cite{Kunz:2005ei,Kunz:2006yp},
giving rise to an instability \cite{Gauntlett:1998fz,Kunz:2005ei}.
For $\lambda > 2 \lambda_{\rm sg}$ uniqueness 
of the black hole solutions is violated
and black holes with a rotating horizon
but with vanishing angular momenta, $J=0$, arise
\cite{Kunz:2005ei}.

To this set of remarkable properties of Einstein-Chern-Simons black holes,
Jose Bl\'azquez-Salcedo added surprizing new results in his talk.
He first analyzed the extremal black hole solutions in 
the near-horizon formalism for $\lambda > 2 \lambda_{\rm sg}$, showing that
for positive charge there are three distinct branches,
while there is a single branch for negative charge.
In an area versus angular momentum diagram, he then pointed out several
particular solutions including the extremal Reissner-Nordstr\"om solution,
a set of two solutions with $J=0$, a set of two
singular solutions with vanishing area, and two more cusp solutions.
Subsequently, he compared with the numerically obtained global solutions,
and showed that they yield a far more intricate phase diagram 
than the near-horizon solutions. 
Consequently, a given near horizon solution
can correspond to either a single global solution, to
more than one global solution, or to no global solution at all
\cite{Blazquez-Salcedo:2013muz,Blazquez-Salcedo:2015kja}.

Moreover, as Jose Bl\'azquez-Salcedo  explained,
there is a whole sequence of extremal rotating global $J=0$ solutions,
and not only two as suggested by the near-horizon formalism.
These solutions possess an increasing number of radial nodes 
in a metric function and a gauge field function,
as he demonstrated for solutions with more than 30 nodes
\cite{Blazquez-Salcedo:2013muz,Blazquez-Salcedo:2015kja}.
Fixing the charge, the mass of this sequence converges 
to the mass of the corresponding extremal Reissner-Nordstr\"om black hole.
He showed, that
these excited extremal black hole solutions are located inside the domain
of existence. Therefore these 
Einstein-Maxwell-Chern-Simons black holes exhibit
a new type of uniquenss violation, namely 
for the same sets of global charges
there can exist both extremal and non-extremal black holes
\cite{Blazquez-Salcedo:2013muz,Blazquez-Salcedo:2015kja}.
Jose Bl\'azquez-Salcedo ended with a glance at his current work,
which considers the presence of a negative cosmological constant.

\subsubsection{$p$-form black holes}

Marcello Ortaggio addressed higherdimensional black holes
in the presence of a $p$-form field instead of the standard electromagnetic
2-form field, while including also a cosmological constant.
He started by explaining the class of four-dimensional exact solutions
given by the Robinson-Trautman family  
\cite{Robinson:1962zz},
which is defined by the existence
of a geodesic and shear-free, twist-free, expanding null vector field. 
This family includes static black holes with a cosmological constant, 
the $C$-metric, radiation, the Vaidya metric, etc.
He then recalled
electrovac Robinson-Trautman spacetimes 
which describe the formation of black holes by gravitational collapse 
of electromagnetic radiation 
\cite{Senovilla:2014nva,Lemos:1997bd,Lemos:1998iy,Dadras:2012yd}.

Turning to Robinson-Trautmann spacetimes in higher dimensions, 
Marcello Ortaggio recalled that the general metric is known
\cite{Podolsky:2006du},
and that the electrovac solutions essentially reduce to
static black holes with an Einstein horizon
\cite{Podolsky:2006du,Ortaggio:2007hs}.
He remarked that the shearfree condition may be
too strong in higher dimensions
\cite{Ortaggio:2012hc,Ortaggio:2012cp,TaghaviChabert:2010bm,TaghaviChabert:2011ex},
and pointed out that in the case of more general $p$-forms 
electromagnetic radiation may have different properties
than in the standard case \cite{Durkee:2010xq,Ortaggio:2014ipa}.

Focussing on the Einstein-Maxwell equations with $p$-forms,
Marcello Ortaggio recalled a number of known results for static $p$-forms
like the no dipole hair theorem \cite{Emparan:2010ni}.
Then he described the construction of 
{\sl Static and Radiating $p$-form Black Holes 
in the Higher Dimensional Robinson-Trautman Class}
\cite{Ortaggio:2014gma}.
He determined the Robinson-Trautman spacetimes in the
presence of a Maxwell $p$-form, but
he restricted to the case of aligned Maxwell fields
\cite{Ortaggio:2014gma}.

Marcello Ortaggio explained that
the properties of these solutions depend strongly on the dimension $D$
and the value of $p$.
In odd dimensions one finds static black holes with an electromagnetic
field \cite{Bardoux:2012aw},
while in even dimensions 
static black holes are also present,
but there are in addition non-static metrics
\cite{Ortaggio:2014gma}.
For $D=2p$, some solutions may describe black hole formation 
by the collapse of electromagnetic radiation \cite{Ortaggio:2014gma}
analogously to \cite{Senovilla:2014nva}.

\subsection{Properties of solutions}

Some of the talks did not present new black hole solutions
but were solely devoted to the discussion of the physical and mathematical 
properties of known higherdimensional black holes,
including their geodesics, stability and greybody factors.

\subsubsection{Geodesics}

For our understanding of the physical properties of black holes 
it is essential to study the motion
of test particles and light in these spacetimes.
As shown in \cite{Kubiznak:2006kt,Page:2006ka,Frolov:2006pe},
the geodesic equations of the higherdimensional Myers-Perry black holes
are separable.
The geodesics in Myers-Perry black hole space-times were studied in
\cite{Frolov:2003en,Gooding:2008tf,Hackmann:2008tu,Kagramanova:2012hw,Diemer:2014lba}.
In contrast, the equations of motion in black ring spacetimes were found to be
separable only in special cases 
\cite{Elvang:2006dd,Hoskisson:2007zk,Durkee:2008an,Igata:2010ye,Armas:2010pw,Igata:2010cd,Grunau:2012ai,Grunau:2012ri,Igata:2013be}.

In the presence of a cosmological constant, the geodesics of
static black holes in four and higher dimensions 
\cite{Hackmann:2008zza,Hackmann:2008tu} and of rotating Kerr-AdS black holes
\cite{Hackmann:2010zz} are known exactly for the general case.
In his talk
{\sl Geodesics of the AdS Myers Perry Black Hole with Equal Angular Momenta}
based on \cite{Delsate:2015ina}
Terence Delsate analyzed the timelike and null geodesics of 
Myers-Perry black holes with AdS asymptotics 
\cite{Hawking:1998kw,Gibbons:2004uw}.
To facilitate the analysis, he restricted to black holes with
equal angular momenta, and he focussed on spherical orbits.

In the discussion of his results, Terence Delsate then addressed
the timelike innermost stable circular orbits and the null circular orbits. 
Interestingly, unlike their asymptotically flat counterparts,
these higherdimensional black holes allow for
bound timelike orbits. 
Terence Delsate then explained that in these AdS spacetimes, 
for sufficiently massive black holes 
there is a parameter range where the spacetime is stable against superradiance
and the innermost stable circular orbit is located inside the ergoregion. 
In contrast, for the massless case, he could not find
stable circular orbits outside the horizon. However, there could be
stable null circular orbits around a naked singularity
\cite{Delsate:2015ina}.

\subsubsection{Greybody factors}

Hawking radiation emitted at the event horizon of a black hole gets modified
by the black hole geometry, yielding a spectrum for an asymptotic observer
which is no longer a black body spectrum, with the difference being encoded in
the greybody factor (see e.g.~ref.~\cite{Kanti:2014vsa} and references therein).
In his talk 
{\sl Greybody Factors of Rotating Cohomogeneity-1 Black Holes}
Ednilton de Oliveira analyzed the greybody factors of
black holes in higher odd dimensions, rotating with equal angular momenta,
for different asymptotics: asymptotically flat black holes, 
de Sitter and anti-de Sitter black holes
\cite{Jorge:2014kra}.
His objectives were to understand how the cosmological constant
influences the greybody factors of rotating black holes
and to analyze superradiance.

Ednilton de Oliveira pointed to the difficulty of this problem,
revealing itself in the scarcity of results on greybody factors 
for rotating and non-asymptotically
flat black holes. For static higherdimensional black holes 
a proper definition of greybody factors for both asymptotically de Sitter
and anti-de Sitter spacetimes was provided in 
\cite{Harmark:2007jy},
where the greybody factors were obtained for scalar fields by the use of
approximate analytic methods, valid in certain regimes of the parameters 
(such as $s$-waves and low frequencies).
Since the methods of \cite{Harmark:2007jy}
can also be applied to more general black hole spacetimes, which
depend on a single (radial) coordinate, 
this suggested the study of the greybody factors of rotating black holes
in odd dimensions with equal angular momenta.

After recalling these cohomogeneity-1 black hole spacetimes,
Ednilton de Oliveira considered a minimally coupled massless scalar field.
Showing that a separation ansatz leads to a simple angular eigenvalue equation,
he concentrated on the radial equation and the effective potential
for the three different asymptotic cases.
He recalled the approximate analytic method \cite{Harmark:2007jy}
and its range of validity, addressed the numerical method employed
in his calculations, and then presented the main results \cite{Jorge:2014kra}.
The analytic results for the $s$-wave modes include the observations
that for asymptotically flat black holes
and low frequencies $\omega$ the greybody factors behave like
$\omega^{D-2}$. For small de Sitter black holes the greybody factor
is given by the ratio of the areas of the black hole and the de Sitter horizon
for small frequencies, while for small anti-de Sitter black holes
the greybody spectrum exhibits a rich structure with large amplitude
oscillations \cite{Jorge:2014kra}.
The numerical calculations showed, that the analytic approximations 
become better with increasing dimension. They also revealed that 
the strongest superradiance effect is seen for $p$-waves,
but decreases with increasing dimension \cite{Jorge:2014kra}.

\section{Solutions in Lovelock gravity}

Lovelock gravity represents a gravity theory
generalizing Einstein's General Relativity to higher dimensions
\cite{Lovelock:1971yv}.
It is based on a symmetric metric tensor endowed with
a Levi-Civita connection. The operator on the r.h.s.~of the field
equations is divergence free, and the field equations are of second order.
While in four dimensions these requirements lead to General Relativity
with a cosmological constant, in higher dimensions
further terms are allowed as shown by Lovelock 
\cite{Lovelock:1971yv}.
In five and higher dimensions, for instance, the
Gauss-Bonnet term can be added,
yielding Einstein-Gauss-Bonnet theory.
Reviews on black holes in Lovelock gravity
can, e.g., be found in refs.~\cite{Charmousis:2008kc,Garraffo:2008hu,Camanho:2011rj}.
Note, that recently arguments were put forward \cite{Camanho:2014apa} showing that
EGB theory is in conflict with causality, unless it has a UV completion 
involving an infinite tower of higher-spin particles with fine-tuned coupling.

\subsection{Black holes in Lovelock gravity}

While general Lovelock theories include all terms allowed for a given
dimension in the action, thus combining the action of General Relativity
with higher order Lovelock terms, in pure Lovelock theory 
the Einstein-Hilbert action is present only in three and four dimensions,
whereas in higher dimensions always only the respective
Lovelock term ${\cal L}_N$, characteristic for the dimension $D$,
where $N=\lfloor (D-1)/2 \rfloor $, together with the
cosmological constant is considered
\cite{Cai:2006pq,Kastor:2012se,Dadhich:2012pd,Dadhich:2015lra}.

\subsubsection{Pure Lovelock black holes}

In his talk on {\sl Gravity in Higher Dimensions}
\cite{Dadhich:2012pd,Dadhich:2015lra}
Naresh Dadhich first explained how
General Relativity would follow from the geometric properties
of Riemann curvature, recalling that 
the Einstein tensor is non-trivial only in $D>2$ dimensions,
while gravity is purely kinematic in $D=3$,
and a non-trivial vacuum solution for free propagation
requires $D > 3$. 

Naresh Dadhich then argued, that in order to
universalize the kinematic property of General Relativity to all odd
dimensions, requiring a generalization of General Relativity
for higher dimensions, pure Lovelock gravity is uniquely singled out.
In his view pure Lovelock gravity is the most natural generalization
of General Relativity because it uniquely
retains the second order field equations, while its
action is a homogeneous polynomial built from the Riemann tensor
\cite{Cai:2006pq,Kastor:2012se,Dadhich:2012pd,Dadhich:2015lra}.
Indeed, in all odd dimensions $D=2N+1$, in pure Lovelock theory
gravity in kinematic \cite{Dadhich:2012cv,Camanho:2015hea}.
He then argued that the pure Lovelock theory based on the $N$-th 
Lovelock term is only valid in $D=2N+1$ odd dimensions
and $D=2N+2$ even dimensions, including General Relativity
for $N=1$.

Turning to black holes, Naresh Dadhich reminded the audience of the
BTZ black hole in three dimensions \cite{Banados:1992wn},
arguing that analogs of BTZ black holes should exist in 
pure Lovelock gravity in all odd dimensions \cite{Dadhich:2012cv}.
Indeed, inspection of the static spherically symmetric black holes
of pure Lovelock theory \cite{Cai:2006pq,Dadhich:2012ma} reveals their presence.
Inspection of the general set of static black hole solutions 
in pure Lovelock gravity shows, that
the gravitational potential in the metric function is essentially the $N$-th root
of the potential in General Relativity, both in odd and even dimensions
\cite{Cai:2006pq,Dadhich:2012eg,Dadhich:2012ma}.
Asymptotically, this yields a $1/r^{D-3}$ dependence of the potential
as for the black holes of General Relativity.
In even dimensions for small values of the radial coordinate 
the potential is of order $1/r^{1/N}$, which allows
for stable bound orbits in pure Lovelock black hole spacetimes
in even dimensions.
Moreover, the characteristic gravitational potential leads to a remarkable
universal thermodynamical behavior in terms of the event horizon radius. 
In particular, the temperature and the entropy always bear
the same relation to the horizon radius in all odd and even dimensions
\cite{Cai:2006pq,Dadhich:2012eg,Dadhich:2012ma}.

\subsubsection{Black holes with nonspherical horizon topology}

In his talk entitled {\sl Static Pure Lovelock Black Hole Solutions 
with Horizon Topology $S(n)\times S(n)$}, 
Josep Pons recalled that in order 
to obtain spherically symmetric static black holes in Lovelock gravity,
one basically has to solve an algebraic polynomial of degree $N$
\cite{Boulware:1985wk,Wheeler:1985qd,Wheeler:1985nh,Whitt:1988ax,Banados:1993ur,Crisostomo:2000bb,Dadhich:2012eg}.
He then argued that the
algebraic character of the ultimate equation holds also for
$S(n)\times S(n)$ black holes,
where the horizon topology is a product of two spheres
\cite{Pons:2014oya,Dadhich:2015nua}.

Starting out with such solutions in General Relativity
Josep Pons recalled the solutions of type
$S(d_0)\times S(d_0)$, $D=2d_0+2$,
reminding the audience of the 
$S^2\times S^2$ merger solution of Kol 
in the context of the nonuniform black string -- caged black hole
transition \cite{Kol:2002xz}.

In Einstein-Gauss-Bonnet gravity 
an interesting example of a black hole solution 
with an $S(d_0)\times S(d_0)$ 
topology was also studied before
\cite{Dotti:2005rc,Dotti:2008pp,Bogdanos:2009pc,Maeda:2010bu,Pons:2014oya}.
Besides having a nontrivial boundary,
the spacetime contains a contribution from the Weyl tensor,
which leads to a slow falloff of the metric function.
Josep Pons explained that the presence of this term
implies that the cosmological constant should always be present 
and positive. Moreover, in addition to the central singularity
there also occurs a noncentral singularity, which may be naked.
The latter can be avoided when the black hole mass and the cosmological
constant satisfy a certain constraint
\cite{Pons:2014oya,Dadhich:2015nua}.

Restricting to the pure Lovelock case,
Josep Pons then considered black holes with
horizon topology $S(N) \times S(N)$ with $D=2N+2$.
In fact, a general pattern appeared in this case,
namely for even $N$, the cosmological constant should be positive,
yielding a black hole horizon and a cosmological horizon
(within a certain range of parameters),
whereas for odd $N$ the cosmological constant should be negative
to obtain a black hole horizon
\cite{Pons:2014oya,Dadhich:2015nua}.
Interestingly, $S(N) \times S(N)$ black holes are
thermodynamically stable for odd $N$ and 
negative cosmological constant, and unstable
for even $N$ and positive cosmological constant.
Moreover, the universal thermodynamical behavior of pure Lovelock black holes
with spherical horizon topology
\cite{Cai:2006pq,Dadhich:2012eg,Dadhich:2012ma}
continues to hold also for these $S(N) \times S(N)$ black holes
\cite{Pons:2014oya,Dadhich:2015nua}.

\subsection{Properties of Lovelock black holes}

In the last part of the session, the properties of known
Lovelock solutions were discussed, in particular,
their thermodynamics and stability.

\subsubsection{Stability of black strings}

Julio Oliva addressed black strings in Lovelock gravity
in his talk entitled
{\sl Black Strings in Gauss-Bonnet Theory are Unstable}
based on \cite{Giacomini:2015dwa}.
He first recalled the analysis of the 
Gregory-Laflamme instability of black strings in
General Relativity in five dimensions
\cite{Gregory:1993vy},
as well as the Gubser-Mitra conjecture,
relating the thermal and perturbative instabilities
of black holes with extended directions
\cite{Gubser:2000mm,Hollands:2012sf}.

Subsequently Julio Oliva turned to black strings
in Einstein-Gauss-Bonnet theory,
and analyzed the behavior of the metric function of the
analytically known black hole solutions
in $D$ dimensions \cite{Boulware:1985wk}.
In particular, he discussed the two limiting cases, where,
depending on the ratio of the radial coordinate and the 
square root of the Gauss-Bonnet coupling constant,
either a Schwarzschild-Tangherlini solution is 
approached or a solution of the pure
Gauss-Bonnet theory (without cosmological constant)
\cite{Crisostomo:2000bb}.

Pointing out that the latter solutions can be oxidated 
to construct homogeneous black string and $p$-brane solutions
\cite{Giribet:2006ec},
Julio Oliva then addressed pure Gauss-Bonnet black strings
(without cosmological constant).
He recalled their thermal instability \cite{Giribet:2006ec},
and argued, that this instability should have a perturbative counterpart.
Proceeding analogously to the Gregory-Laflamme instability analysis
in General Relativity,
he then analyzed the perturbative instability of pure
Gauss-Bonnet black strings in seven dimensions,
and came to the analogous picture concerning the instability
of the black string in pure Gauss-Bonnet theory 
as the one known from General Relativity
\cite{Giacomini:2015dwa}.

\subsubsection{Thermodynamics}

In her talk
{\sl Lovelock Black Hole Thermodynamics}
Antonia Frassino addressed
phase transitions of
static spherically symmetric and asymptotically AdS
black holes in Maxwell-Lovelock gravity,
focussing on 2nd and 3rd oder Lovelock terms
\cite{Frassino:2014pha}
(for earlier work see e.g.~\cite{Cai:1998vy,Cai:2001dz,Cai:2003kt,Cai:2003gr,Kim:2007iw}).
She first recalled that
by identifying the negative cosmological constant 
with a thermodynamical pressure,
and introducing an associated conjugate thermodynamic volume
one obtains a generalized first law and Smarr formula
both in General Relativity \cite{Kastor:2009wy}
and in Lovelock gravity \cite{Kastor:2010gq,Kastor:2011qp}.

In her thermodynamical investigations Antonia Frassino
then kept the Lovelock coupling constants fixed except for
the cosmological constant and studied the possible
phase transitions based on the behavior of the Gibbs free energy
in the canonical ensemble. 
She pointed out that the pressure must be sufficiently small
in these solutions, since otherwise the solutions 
would not possess an asymptotic AdS region,
and space would become compact.

While previous studies 
\cite{Zou:2014mha,Wei:2014hba,Mo:2014mba,Wei:2012ui,Cai:2013qga,Xu:2013zea}
of the Gauss-Bonnet (2nd order Lovelock) case had
already shown a van der Waals behavior \cite{Kubiznak:2012wp},
as well as reentrant phase transitions and tricriticality,
Antonia Frassino added 
that a triple point arises only in six dimensions and has
no counterpart in higher dimensions
\cite{Frassino:2014pha}.
She then explained her new results for
the 3rd order Lovelock case, 
where for hyperbolic Lovelock black holes
multiple reentrant phase transistions occur,
and for special tuned Lovelock couplings a new
type of isolated critical point appears
\cite{Frassino:2014pha}.

\section{Conclusions}

The field of higherdimensional black holes has seen
most interesting developments during the last years.
Besides much progress with analytical methods
also major progress concerning numerical methods has been achieved.
Here in particular the highly sophisticated pseudospectral
methods promise to yield further impressive results
in the near future, which will hopefully be reported in this session at MG15.

\end{document}